\documentclass[10pt,aps,prd,twocolumn,showpacs,amsmath,amssymb,nofootinbib,eqsecnum,preprintnumbers,longbibliography,superscriptaddress]{revtex4-1}

\usepackage[english]{babel}				
\usepackage[mathscr]{eucal} 			
\usepackage{accents}					
\usepackage{mathtools}					
\usepackage[ddmmyyyy]{datetime}    		

\usepackage{graphicx}
\usepackage{xcolor}
\usepackage{float}

\newcommand\feq{\mathrel{\phantom{=}}}

\DeclareMathOperator{\sh}{sh}

\let\th\relax
\DeclareMathOperator{\th}{th}

\DeclareMathOperator{\erfc}{erfc}

\begin{document}
	
\title{Impulsive waves in ghost free infinite derivative gravity in anti-de~Sitter spacetime}

\author{Suat Dengiz}

\email{sdengiz@thk.edu.tr}

\affiliation{Department of Mechanical Engineering, University of Turkish Aeronautical Association, 06790 Ankara, Turkey}

\author{Ercan Kilicarslan}

\email{ercan.kilicarslan@usak.edu.tr}

\affiliation{Department of Physics, Usak University, 64200, Usak, Turkey.}

\author{Ivan  Kol\'a\v{r}}	

\email{i.kolar@rug.nl}
	
\affiliation{Van Swinderen Institute, University of Groningen, 9747 AG, Groningen, The Netherlands.}

\author{Anupam Mazumdar}
	
\email{anupam.mazumdar@rug.nl}
	
\affiliation{Van Swinderen Institute, University of Groningen, 9747 AG, Groningen, The Netherlands.}

\date{\today}
	
\begin{abstract} 
We study exact impulsive gravitational waves propagating in anti-de Sitter spacetime in the context of the ghost free infinite derivative gravity. We show that the source-free theory does not admit any AdS wave solutions other than that of Einstein's general relativity. The situation is significantly different in the presence of sources. We construct impulsive-wave solutions of the infinite derivative gravity generated by massless particles and linear sources in four and three dimensions. The singularities corresponding to distributional curvature at the locations of the sources get smeared by the non-localities. The obtained solutions are regular everywhere. They reduce to the corresponding solutions of general relativity in the infrared regime and in the local limit.
\end{abstract}

\maketitle
\section{Introduction}
Einstein's general relativity (GR) has surpassed all observations from solar system tests to gravitational waves so far \cite{will}. However, it is not well constrained at short distances, i.e., in the ultraviolet regime (UV). Newton's $1/r$-potential was experimentally tested up to approximately $5$~$\mu$m \cite{Kapner}, which corresponds to ${0.001}$~eV. Beyond these scales, gravitational interaction has not been constrained by direct experiments. Furthermore, as one approaches the short distances, GR has several problems. From the classical point of view, it suffers from the presence of spacetime singularities \cite{Hawking}; at the quantum level, it fails to be perturbatively renormalizable.

It has been known for a while that non-local terms actions can improves UV behavior. Non-local theories containing \textit{form factors} with an infinite number of derivatives have brought considerable interest in the context of quantum field theories \cite{Efimov:1967pjn, Efimov:1971wx, Efimov:1972wj, Krasnikov:1987yj, Moffat:1990jj, Evens:1990wf, Tomboulis:2015gfa, Buoninfante:2018mre} and quantum gravity \cite{Tomboulis:1980bs,Tomboulis:1997gg,Modesto1,Tseytlin:1995uq,Siegel:2003vt}. In particular, it was shown that infinite derivative gravity (IDG) may resolve cosmological \cite{Biswas2} and black hole singularities \cite{Biswas1}. In order to avoid introducing ghost-like instabilities, the form factors are chosen as analytic functions with no roots in the complex plane (i.e., exponential of entire functions), see~\cite{Tomboulis:1980bs,Tomboulis:1997gg,Biswas1}. Moreover, the form factor of such a non-local action emerges from the world line approximation of one-loop amplitude in string theory~\cite{Abel:2019ufz,Abel:2019zou}. There were also first attempts in studying initial value problem of IDG using diffusion equation method \cite{Calcagni:2018gke,Calcagni:2018lyd} and constructing perturbative Hamiltonian \cite{Kolar:2020ezu} using non-local Hamiltonian formalism of \cite{Llosa:1993sj,Gomis:2000gy}.

Recently, there has been further progress in finding solutions of linearized IDG. It was shown that IDG may avoid not only black-hole type singularities \cite{Frolov:2015bia,Frolov:2015usa,Edholm:2016hbt,Kilicarslan:2018yxd,Buoninfante:2018stt,Buoninfante:2018xiw,Buoninfante:2018xif,Boos:2020ccj}, but also topological defects such as p-branes \cite{Boos:2018bxf}, cosmic strings \cite{Boos:2020kgj}, and NUT-like singularities \cite{Kolar:2020bpo}. The exact pp-wave solutions have been studied in \cite{Kilicarslan:2019njc}. 

In this paper, we study the non-expanding gravitational waves of the Siklos type in anti-de Sitter universe, the \textit{AdS waves}, which are generalizations of the so-called pp-waves in flat space in the context of the ghost-free infinite derivative gravity presented in \cite{Biswas:2016etb,Biswas:2016egy}. The main focus of this work are the \textit{impulsive waves}, which have been studied extensively in GR with a cosmological constant \cite{Hotta:1992qy, Podolsky1, Podolsky2, Podolsky3, Podolsky4, Podolsky5, Cai:1999dz}. These solutions are generated by null sources with Dirac-delta stress-energy tensor and belong to the class of almost universal spacetimes \cite{Hervik:2013cla,Kuchynka:2018ezs}. The impulsive-wave solution of IDG corresponding to a massless point particle was obtained in \cite{Kilicarslan:2019njc}. Here, we follow up by extending the analysis to the AdS spacetime in four and three dimensions. We illustrate how the non-locality affects the gravitational waves in AdS if the sources are absent or present.

The lay-out of the paper is as follows: In Sec.~\ref{sc:IDG}, we briefly review the ghost-free infinite derivative gravity. In Sec.~\ref{sc:APW}, we study the AdS wave solutions in the source-free case. Sec.~\ref{sc:SOL4D} and Sec.~\ref{sc:SOL3D} are dedicated to the constructions of the impulsive gravitational waves of IDG in ${3+1}$ and ${2+1}$ dimensions, respectively. In Sec.~\ref{sc:CON}, we conclude with a brief discussion of our results. Supplementary material is attached to the appendices.
\section{Infinite Derivative Gravity} \label{sc:IDG}
The most general quadratic in curvature (parity-invariant, and torsion-free) theory of IDG in four dimensions with a cosmological constant $\Lambda$, \cite{Biswas2,Biswas1,Biswas:2016etb,Biswas:2016egy} is given by the Lagrangian density\footnote{We use mostly positive metric signature, $(-,+,+,+)$.}
\begin{equation}\label{action}
\begin{aligned}
\mathcal{L} &= \frac{\sqrt{-g}}{16\pi G}\Big[ R-2\Lambda +\alpha_c\big(R {\cal
	F}_1 (\square_s) R +  R_{\mu\nu} {\cal
	F}_2(\square_s) R^{\mu\nu}
	\\
&\feq+ C_{\mu\nu\rho\sigma} {\cal
	F}_3(\square_s) C^{\mu\nu\rho\sigma}\big)\Big],
\end{aligned}
\end{equation}
where ${G=M^{-2}_p}$ is Newton's gravitational constant, ${\square_s\equiv\square/M_s^{2}}$, and $\alpha_c={1}/{M_s^2}$. The dimensionful constant $M_s$ is the \textit{scale of non-locality} at which non-local interactions become manifest. In the \textit{local limit}, ${M_s \to \infty}$, the theory reproduces Einstein's general relativity. The form factors ${\cal F}_i(\square_s)$ are analytic functions of d'Alembert operator
${\square\equiv g_{\mu\nu}\nabla^\mu\nabla^\nu}$,
\begin{equation}
{\cal F}_i(\square_s)\equiv\sum_{n=0}^{\infty}f_{i,n}\frac{\square^n}{M_s^{2n}}, 
\label{idfunc}
\end{equation} 
where $f_{i,n}$ are dimensionless coefficients. The form factors give rise to non-local gravitational interactions. They are crucial to make the theory ghost-free, and the analyticity is required for obtaining the low energy limit similar to that of GR.
 The equations of motion for the action \eqref{action} are given in Appendix~\ref{ap:EOM}.
 
\section{AdS wave spacetimes in IDG}\label{sc:APW}

The field equations of the infinite derivative gravity are very complicated \cite{Biswas:2013cha}, so a mere attempt of finding exact solutions to the theory is an extremely daunting task. To handle the situation, we focus on the AdS wave metric ansatz, which can be written in the Kerr-Schild form,\footnote{For detailed properties of AdS waves and the Kerr-Schild metrics, we refer the reader to \cite{Kerr-Schild,Classification,Gullu:2011sj,Gurses:2012db}. }
\begin{equation}
g_{\mu \nu}=\bar{g}_{\mu \nu}+2 H \lambda_\mu \lambda_\nu,
\label{KSForm}
\end{equation}
where $\bar{g}_{\mu \nu}$ denotes the AdS background metric, and  $H$ is a scalar function that satisfies $\lambda^\mu\partial_\mu H=0$. Here, $\lambda_\mu$ is a non-expanding, non-twisting, and shear-free null vector satisfying
\begin{equation}
\begin{aligned}
\lambda^{\mu}\lambda_\mu=0,\quad\nabla_{\mu}\lambda_\nu=\xi_{(\mu}\lambda_{\nu)},\quad\xi_{\mu}\lambda^\mu=0,
\label{AdSplanewave}
\end{aligned}
\end{equation}
where $\xi_{\mu}$ is a vector in the transverse direction. Due to the fact that the curvature scalar $R$ is constant, there is no contribution from the non-local form factor $R{\cal F}_{1}(\square_s)R$ to the field equations except a constant term. In addition, the Ricci tensor becomes \cite{Gursespp,Gurses:2012db,Gurses:2013jua}
\begin{equation}
R_{\mu\nu}=-\frac{3}{\ell^2}g_{\mu\nu}+\lambda_{\mu}\lambda_{\nu}\mathcal{O}H,
\label{Ricci}
\end{equation}
where $\ell$ is the AdS radius and $\mathcal{O}$ denotes the operator
\begin{equation}
\mathcal{O}\equiv-\left(\square+2\xi^{\mu}\partial_{\mu}+\frac{1}{2}\xi^{\mu}\xi_{\mu}-\frac{4}{\ell^{2}}\right).
\end{equation}
Furthermore, one should note that the traceless Ricci tensor takes the form
\begin{equation}
S_{\mu\nu}=\lambda_{\mu}\lambda_{\nu}\mathcal{O}H,
\label{TraclessRicci}
\end{equation} 
which is of the type N in the aspect of null alignment classification \cite{Coley1,Coley2}. Moreover, one can derive the following formulas for the (repeated) action of the d'Alembert operator \cite{Gursespp}:
\begin{equation}\label{eq:formulas}
\begin{aligned}
&\square(\lambda_\mu\lambda_\nu H) =\bar{\square}(\lambda_\mu\lambda_\nu H)=-\lambda_\mu\lambda_\nu\bigg(\mathcal{O}+\frac{2}{\ell^2}\bigg)H
\\
&\square^{n}S_{\mu\nu} =\bar{\square}^{n}S_{\mu\nu}=\left(-1\right)^{n}\lambda_{\mu}\lambda_{\nu}\left(\mathcal{O}+\frac{2}{\ell^{2}}\right)^{\!\!n}\!\mathcal{O}H,
\end{aligned}
\end{equation}
where $\bar{\square}=\bar{g}^{\mu\nu}\bar{\nabla}_\mu\bar{\nabla}_{\nu}$ is the AdS background d'Alembert operator. Throughout the calculations, one needs to use the following identity of higher-order derivative of the Weyl tensor:
\begin{equation}
\nabla_\mu\nabla_\nu\square^{n}C^{\mu\alpha\nu\beta}=\frac{1}{2}\Big(\square+\frac{R}{3}\Big)^{\!n}\Big(\square-\frac{R}{3}\Big)S^{\alpha\beta}.
\end{equation}

By using the recursive relations above, one can easily convert the field equations of the IDG for the AdS wave metric to a rather simple form,
\begin{equation}
\begin{aligned}
\bigg(\Lambda {+}\frac{3}{\ell^2}\bigg)g_{\mu\nu}{+}\bigg[1{+}\alpha_c\Big[\Big(2f_{1,0}{+}\frac{f_{2,0}}{2}\Big)R{+}\Big(\bar\square{+}\frac{2}{\ell^2}  \Big){\cal F}_{2}(\bar\square_s)
\\
+2{\cal F}_{3}\Big(\bar\square_s{-}\frac{4}{M_s^2\ell^2}\Big)\Big(\bar\square+\frac{4}{\ell^2}\Big)\Big]\bigg]S_{\mu\nu}=0.
\label{ppgeneq1}
\end{aligned}
\end{equation}
The trace part of the equation determines the cosmological constant in terms of the AdS radius:
 \begin{equation}
 \Lambda=-\frac{3}{\ell^2}.
 \end{equation}
Note that \eqref{ppgeneq1} reduces to the field equations for pp-waves on Minkowski background \cite{Kilicarslan:2019njc} in the limit ${\ell\to\infty}$ (i.e., ${\Lambda\to0}$). The traceless part of the field equations yields non-local equations
\begin{equation}
\begin{aligned}
&\bigg[1+\alpha_c\Big[-\frac{12}{\ell^2}\Big(2f_{1,0}+\frac{f_{2,0}}{2}\Big)+\Big(\bar{\square}+\frac{2}{\ell^2} \Big){\cal F}_{2}(\bar{\square}_s)\\
&+2{\cal F}_{3}\bigg(\bar{\square}_s-\frac{4}{M_s^2\ell^2}\bigg)\Big(\bar{\square}+\frac{4}{\ell^2}\Big)\Big]\bigg]\bigg(\bar{\square}+\frac{2}{\ell^2}\bigg)\lambda_\mu\lambda_\nu H=0.
\label{AdSppgeneq}
\end{aligned}
\end{equation}
It is important to stress here that the full equations for AdS waves \eqref{AdSppgeneq} are equivalent to the linearized field equations for the Kerr-Schild perturbations ${h_{\mu\nu}=g_{\mu\nu}-\bar{g}_{\mu\nu}=2H\lambda_{\mu}\lambda_{\nu}}$. Therefore, the solutions of the full equations that we obtain below are also solutions of the linearized equations for the transverse-traceless fluctuations around AdS background.

To ensure that the theory has no extra degrees of freedom and no ghosts on the AdS background, we choose the form factors \footnote{Let us remark that this choice of ${\cal F}_{3}(\square_s)$ is non-analytic. An alternative analytic choice ${\cal F}_{3}(\square_s) =\frac{1}{2}\Big(e^{- (\square_s+\frac{8}{\ell^2M_s^2})}-1\Big)/(\square_s+\frac{8}{\ell^2M_s^2})$ (discussed in \cite{SravanKumar:2019eqt}) would only affect the overall constant $L$ and $L_4$ by the factor $e^{2/\ell^2M_s^2}$ without changing our conclusions.} \cite{Biswas:2016egy}:
\begin{equation}
\begin{gathered}
{\cal F}_{1}(\square_s)={\cal F}_{2}(\square_s)=0,
\\
{\cal F}_{3}(\square_s) =\frac{1}{2}\frac{e^{- (\square_s+\frac{6}{\ell^2M_s^2})}-1}{\square_s+\frac{8}{\ell^2M_s^2}}.
\end{gathered}
\end{equation}
The AdS wave equation \eqref{AdSppgeneq} then turns into
\begin{equation}
e^{-(\bar{\square}_s+\frac{2}{M_s^2\ell^2})} \bigg(\bar{\square}+\frac{2}{\ell^2}\bigg)\lambda_\mu\lambda_\nu H=0.
\label{IDGfeq}
\end{equation}
Let us write AdS wave metric \cite{Siklos} using the null coordinates, ${u=(x-t)/\sqrt{2}}$ and ${v=(x+t)/\sqrt{2}}$,
\begin{equation}
ds^{2}=\frac{\ell^2}{z^2}\big( 2dudv+dy^2+dz^2\big)+2H(u,y,z)du^{2},
\label{AdS-ppmetric}
\end{equation}
where $z=0$ corresponds to the conformal infinity of AdS spacetime \cite{Podolsky:1997ik}. In these coordinates, ${\xi_\mu=2z^{-1}\delta^z_{\mu}}$, and, thus,
\begin{equation}
\mathcal{O}=-\left(\bar{\square}+\frac{4z}{\ell^{2}}\partial_{z}-\frac{2}{\ell^{2}}\right), \,\,\,\,\, \bar{\square}=\frac{z^2}{\ell^2}\partial^2-\frac{2z}{\ell^2}\partial_z-\frac{4z^2}{\ell^2}\partial_u\partial_v,
\label{lightcnuv}
\end{equation}
where we introduced ${\partial^2\equiv\partial_y^2+\partial_z^2}$.

Employing the first formula of \eqref{eq:formulas}, the field equations \eqref{IDGfeq} reduce to
\begin{equation}\label{homogeq}
e^{-\frac{z^2\partial^2+2z\partial_{z}-2}{M_s^2\ell^2}} \left(z^2\partial^2+2z\partial_{z}-2\right)H=0.
\end{equation}
This equation can be solved using the \textit{eigenvalue method} described in \cite{BarnabyKamran}. Let us consider the eigenvalue problem of the operator in the round brackets,
\begin{equation}
	\left(z^2\partial^2+2z\partial_{z}-2\right)H_w=-w^2 H_w,
\end{equation}
where $H_w$ are eigenfunctions and $w$ are the corresponding eigenvalues. By acting with the full non-local operator on $H_w$, we obtain
\begin{equation}
\begin{aligned}
e^{-\frac{z^2\partial^2+2z\,\partial_{z}-2}{M_s^2\ell^2}}\left(z^2\partial^2{+}2z\partial_{z}{-}2\right)H_w
=-e^{\frac{w^2}{M_s^2\ell^2}}w^2 H_w.
\end{aligned}
\end{equation}
The general solution $H$ of the linear equation \eqref{homogeq} is a superposition of such functions $H_w$ for which ${e^{w^2/M_s^2\ell^2}w^2=0}$. Since the exponential has no roots in the complex plane, the only solution is the function ${H_0}$ (the eigenvalue ${w=0}$). Therefore, the original equation effectively reduces just to the equation
\begin{equation}\label{greq}
\left(z^2\partial^2+2z\partial_{z}-2\right)H=0.
\end{equation}
In other words, the only AdS wave solutions of the source-free theory are those of the Einstein's general relativity.\footnote{This is true also in the Minkowski background, which was studied in \cite{Kilicarslan:2019njc}. The source-free solution presented in \cite{Kilicarslan:2019njc} is incorrect because of a mistake in the Fourier transform.} The solutions of \eqref{greq} are well known \cite{Chamblin:1999cj,Gullu:2011sj},
\begin{equation}
H(u,y,z)=z^{-\frac{1}{2}}\big[c_1I_{\frac{3}{2}}(\zeta z)+c_2 K_{\frac{3}{2}}(\zeta z)\big]\sin(\zeta y+c_3),
\label{GRsol}
\end{equation}
where, $\zeta$ and $c_i$ are functions of the null coordinate $u$. The functions $I_{3/2}$ and $K_{3/2}$ are modified Bessel functions of the first and second kind, respectively.

In fact, this result is expected since the source-free theory is not affected by non-localities if the equations of motion are linear, which is exactly the case of the field equations with the AdS wave metric ansatz. In order to see the non-local effects, we need to consider the field equations with a non-zero source. The equations derived above remain intact in the presence of non-zero null sources $T_{\mu\nu}dx^\mu dx^\nu=T_{uu}du^2$ \footnote{A feasible way of such a source is to consider a non-minimally coupled scalar field with a certain potential \cite{AyonBeato:2005,AyonBeato:2006}.}.

\section{Impulsive waves in 3+1 dimensions}\label{sc:SOL4D}

In this section, we will search for impulsive gravitational waves\footnote{For the details on the impulsive gravitational waves in GR with a cosmological constant, see \cite{Hotta:1992qy, Podolsky1, Podolsky2, Podolsky3, Podolsky4, Podolsky5, Cai:1999dz}.} that are generated by massless sources in IDG. Since we put a non-zero stress-energy on the right-hand side of equations of motion, we can expect that the resulting solutions will be affected by the presence of non-local form-factors with infinite derivatives.

\subsection{Massless point-like source}

Let us begin with the impulsive AdS wave metric,
\begin{equation}
ds^2=\frac{\ell^2}{z^2}\big(2du dv+dy^2+dz^2\big)+2\delta(u)H(y,z) du^2,
\label{ansatz}
\end{equation}
and consider a massless point particle traveling in the positive $x$-direction with momentum ${p^\mu=E(\delta^\mu_t+\delta^\mu_x)}$. Such a particle is described by a source with stress-energy tensor ${T_{uu}=E z_0^2\ell^{-2} \delta(u)\delta(y)\delta(z{-}z_0)}$. The AdS wave equation then reads
\begin{equation}
e^{-\frac{z^2\partial^2+2z\partial_{z}-2}{M_s^2\ell^2}} \left(z^2\partial^2{+}2z\partial_{z}{-}2\right)H(y,z)
={-} L \delta(y)\delta(z{-}z_0),
\label{4Deq}
\end{equation}
where we introduced the constant ${L=16\pi G E z_0^2}$. Let us recall that the homogeneous solution is given by \eqref{GRsol}. Since it is the same for the local as well as non-local theory, we will focus on finding a particular solution only.

\begin{figure}[htbp]
	\centering
	\includegraphics[width=\columnwidth]{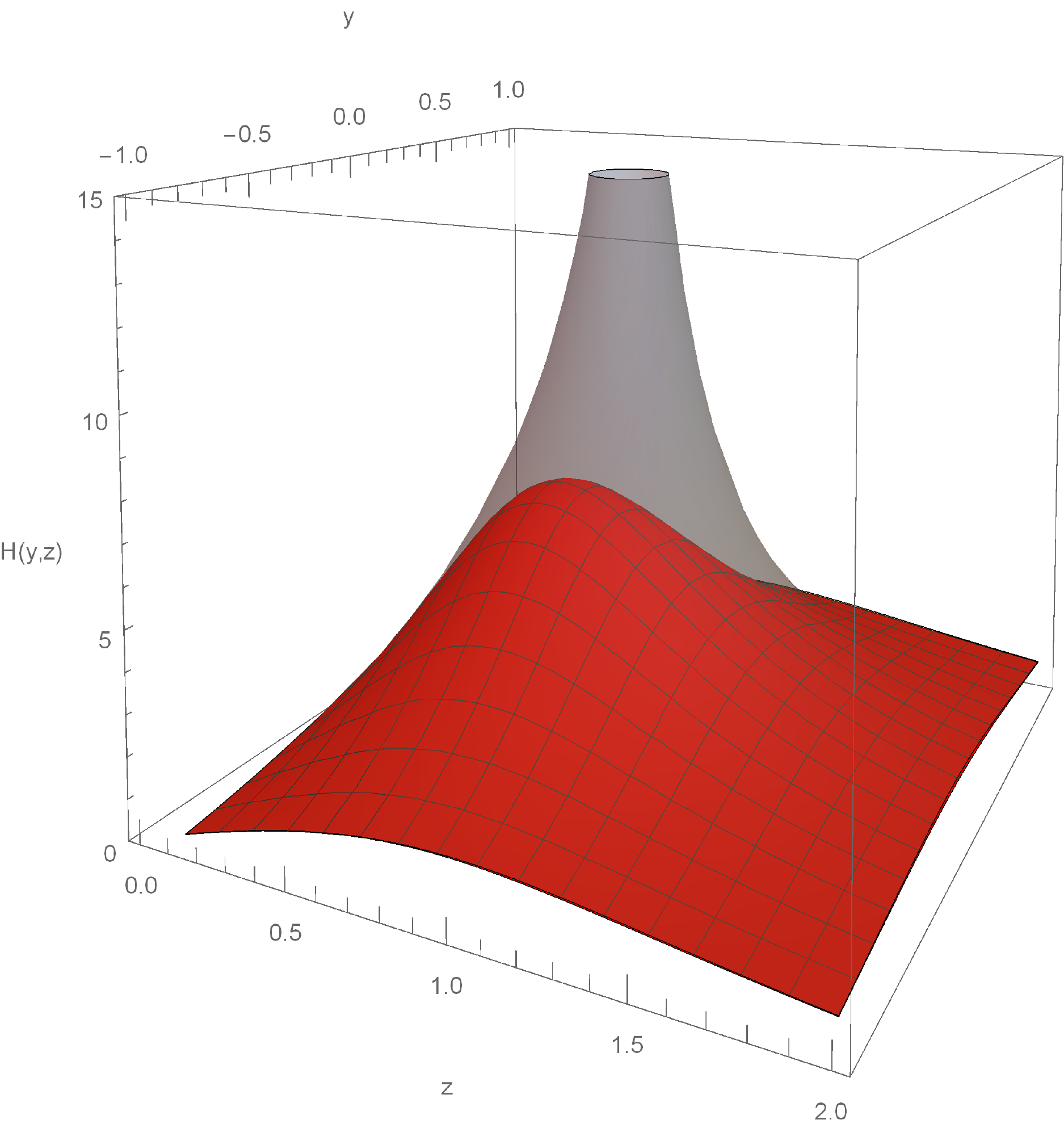}
	\caption{The function $H(y,z)$ for ${z_0=1}$, ${G=1}$, ${E=1}$, ${\ell=1}$, and ${M_s=4}$. The meshed red surface represents the solution of IDG and the grey surface depicts the corresponding solution of GR.}
	\label{GraphSol1}
\end{figure}

In order to solve \eqref{4Deq}, we first take the Fourier transform in coordinate $y$,\footnote{
Our convention for the Fourier transform is:
\begin{equation*}
\begin{aligned}
\hat{f}(k) =\frac{1}{\sqrt{2\pi}}\!\!\int_{\mathbb{R}}\!\!\!d x\,f(y)e^{-i ky}\;,
\quad
f(y) =\frac{1}{\sqrt{2\pi}}\!\!\int_{\mathbb{R}}\!\!\!d \mathrm{k}\,\hat{f}(k) e^{i k y}.
\end{aligned}
\end{equation*}
}
\begin{equation}
\begin{aligned}
e^{-\frac{z^2\partial_z^2+2z\partial_{z}-k^2z^2-2}{M_s^2\ell^2}} \left(z^2\partial_z^2{+}2z\partial_{z}{-}k^2z^2{-}2\right)\hat{H}(k,z)
\\
=-\frac{L}{\sqrt{2\pi}} \delta(z-z_0).
\label{4Deq1}
\end{aligned}
\end{equation}
Using the substitution ${\hat{H}(k,z)=V(k,z)/\sqrt{z}}$, we can rewrite this equation as
\begin{equation}\label{eq:eAk}
e^{-\mathcal{A}(k)/M_s^2\ell^2}\mathcal{A}(k)V(k,z)=-\frac{L\sqrt{z_0}}{\sqrt{2\pi}}\delta(z-z_0)\;,
\end{equation}
where we introduced the $k$-dependent operator
\begin{equation}
\mathcal{A}(k)\equiv z^2\partial_z^2{+}2z\partial_{z}{-}k^2z^2{-}2\;.
\end{equation}
Similar to the homogeneous case, we will first study the eigenvalue problem for this operator. Assuming ${k>0}$, one can show that
\begin{equation}
\mathcal{A}(k)K_{i\beta}(kz)=-(\beta^2+9/4)K_{i\beta}(kz)\;,
\end{equation}
where $K_{i\beta}$ are modified Bessel functions of imaginary order. In order to make further progress, it is essential to express the right-hand side of \eqref{eq:eAk} in terms of the eigenfunctions $K_{i\beta}(kz)$. Fortunately, this is possible thanks to the identity presented in \cite{MurashimaKiyono},
\begin{equation}
\delta(z-z_0)=\frac{2}{\pi^2z_0}\!\int_0^\infty \!\!\!d\beta\, \beta \sh(\pi\beta) K_{i\beta}(kz_0)K_{i\beta}(kz)\;,
\end{equation}
for arbitrary ${k>0}$.
Thus, we can write
\begin{equation}
\begin{aligned}
V(k,z) &=-\frac{L\sqrt{z_0}}{\sqrt{2\pi}}\frac{e^{\mathcal{A}(k)/M_s^2\ell^2}}{\mathcal{A}(k)}\delta(z-z_0)
\\
&=\frac{\sqrt{2}L}{\pi^{\frac52}\sqrt{z_0}}\!\int_{0}^{\infty}\!\!d\beta\,\frac{e^{-(\beta^2+9/4)/M_s^2\ell^2}}{\beta^2+9/4}\beta\sh(\pi\beta)
\\
&\feq\times K_{i\beta}(kz_0)K_{i\beta}(kz)\;.
\end{aligned}
\end{equation}
After taking the inverse Fourier transform, the particular solution of \eqref{4Deq} takes the form of the integral
\begin{equation}\label{eq:solHyz}
\begin{aligned}
	H(y,z)\! &=\!\frac{16GE z_0^{\frac{3}{2}}}{\pi^2\sqrt{z} }\!\!\int_\mathbb{R}\!\! dk\!\int_{0}^{\infty}\!\!\!\! d\beta\, \frac{e^{-(\beta^2{+}9/4)/M_s^2\ell^2}}{\beta^2+9/4}\beta\sh(\pi\beta)
	\\
	&\feq\times K_{i\beta}(|k|z_0)K_{i\beta}(|k|z)e^{iky},
\end{aligned}
\end{equation}
where we also employed the fact that $H(y,z)=H(-y,z)$, as it follows from \eqref{4Deq}. This integral does not seem to have a closed form, but we can evaluate it numerically as shown in Fig.~\ref{GraphSol1}.

The GR solution can be obtained by taking the local limit ${M_s\to\infty}$ of the integrand in \eqref{eq:solHyz}. Using the identity \cite{Ryzhik},
\begin{equation}
\int_{0}^{\infty}\!\!\!\! d\beta\, \frac{\beta\sh(\pi\beta)}{\beta^2+9/4} K_{i\beta}(|k|z_0)K_{i\beta} =
\begin{cases}
\frac{\pi^2}{2}I_{\frac32}(|k|z)K_{\frac32}(|k|z_0),
\\
\frac{\pi^2}{2}I_{\frac32}(|k|z_0)K_{\frac32}(|k|z),
\end{cases}\!\!\!\!
\end{equation}
which holds for ${z<z_0}$ and ${z>z_0}$, respectively, we arrive at the function
\begin{equation}\label{eq:GR4d}
H_{\textrm{GR}}\!=\!\frac{2GE}{z^2}\Big[(y^2{+}z^2{+}z_0^2)\log\Big(1+\frac{4zz_0}{y^2{+}(z{-}z_0)^2}\Big)-4zz_0\Big].
\end{equation}
This GR solution represents an impulsive gravitational wave that is generated by a massless particle, see, for example, \cite{Hotta:1992qy, Podolsky2, Camanho:2014apa}.

It is clear that the impulsive-wave solution of GR diverges at the location of the particle, where it has distributional curvature. On the other hand, the non-local impulsive-wave solution of IDG is regular everywhere due to the improved behavior of the propagator in the UV scale. Let us remark that we could replace $\delta(u)$ by a more realistic smooth regularization of Dirac-delta ${\delta_\epsilon(u)}$ thanks to the linearity of equations and the independence of the coordinate $v$ (derivative $\partial_u$ in \eqref{lightcnuv} never applies). In this sense, all curvature tensors can be considered as regular. Near the conformal infinity ${z=0}$, the non-local solution approaches GR.

\subsection{Massless linear source}

Let us consider a specific example of a null matter distribution, ${T_{uu}=E z_0\ell^{-2}\delta(u)\delta(z{-}z_0)}$, for which one can find an impulsive-wave solution in a closed form. This particular stress-energy tensor describes a linear null source that moves in $x$-direction with momentum ${p^\mu=E(\delta^\mu_t+\delta^\mu_x)}$ and extends to infinity in $y$-direction. The trajectory of this surface is visualized in the Poincar\'e spherical model of Lobachevsky space in Fig.~\ref{lob}. Details of this representation are reviewed in the Appendix~\ref{ap:PSM}.

\begin{figure}[htbp]
	\centering
	\includegraphics[width=0.8\columnwidth]{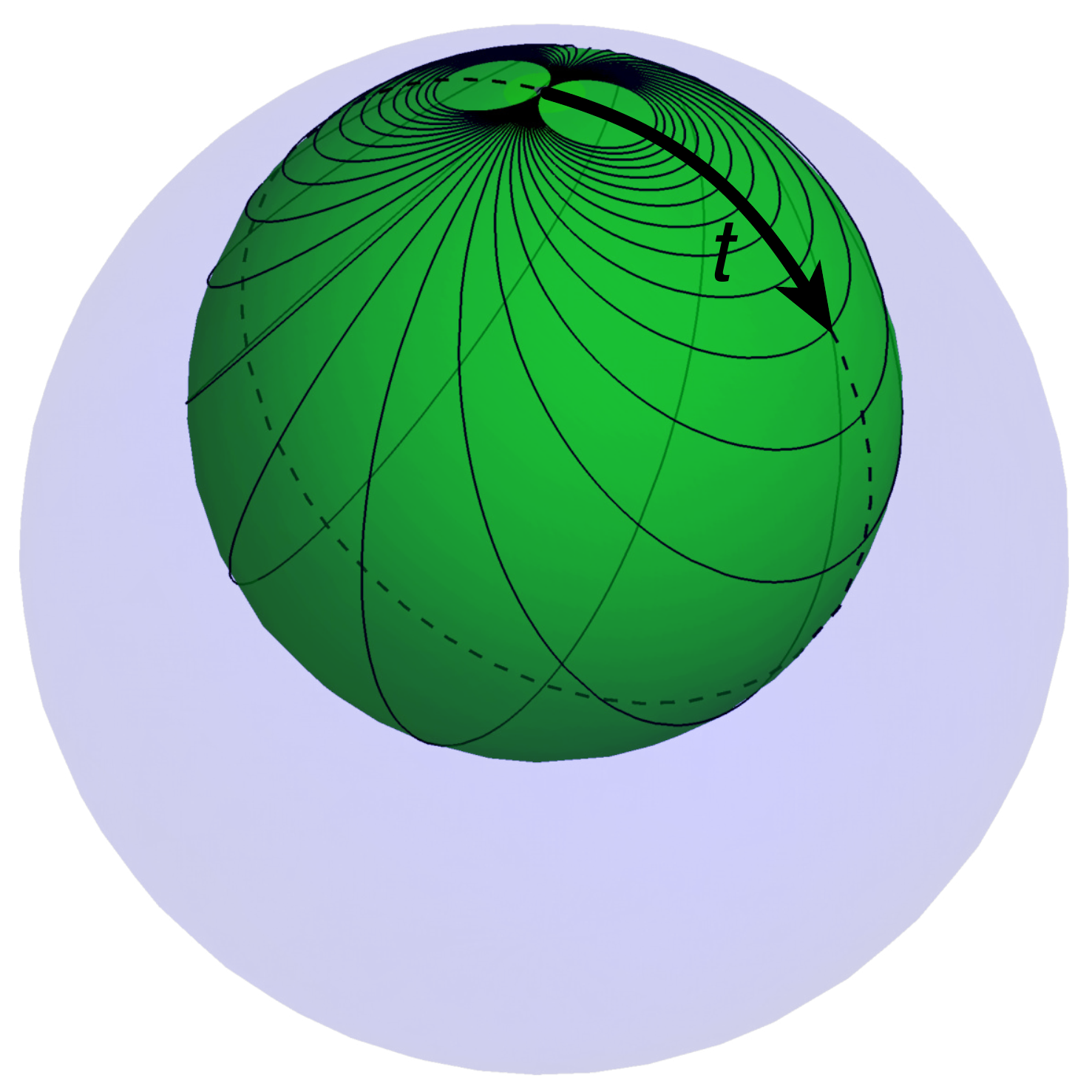}
	\caption{The trajectory of the source at ${z=z_0}$ represented in Poincar\'e spherical model of the Lobachevsky space. The solid lines correspond to the location of the source at a given time $t$. They extend from ${y=0}$ (at the dashed line) toward ${y=\pm\infty}$ (at the conformal infinity).}
	\label{lob}
\end{figure}

\begin{figure}[htbp]
	\centering
	\includegraphics[width=\columnwidth]{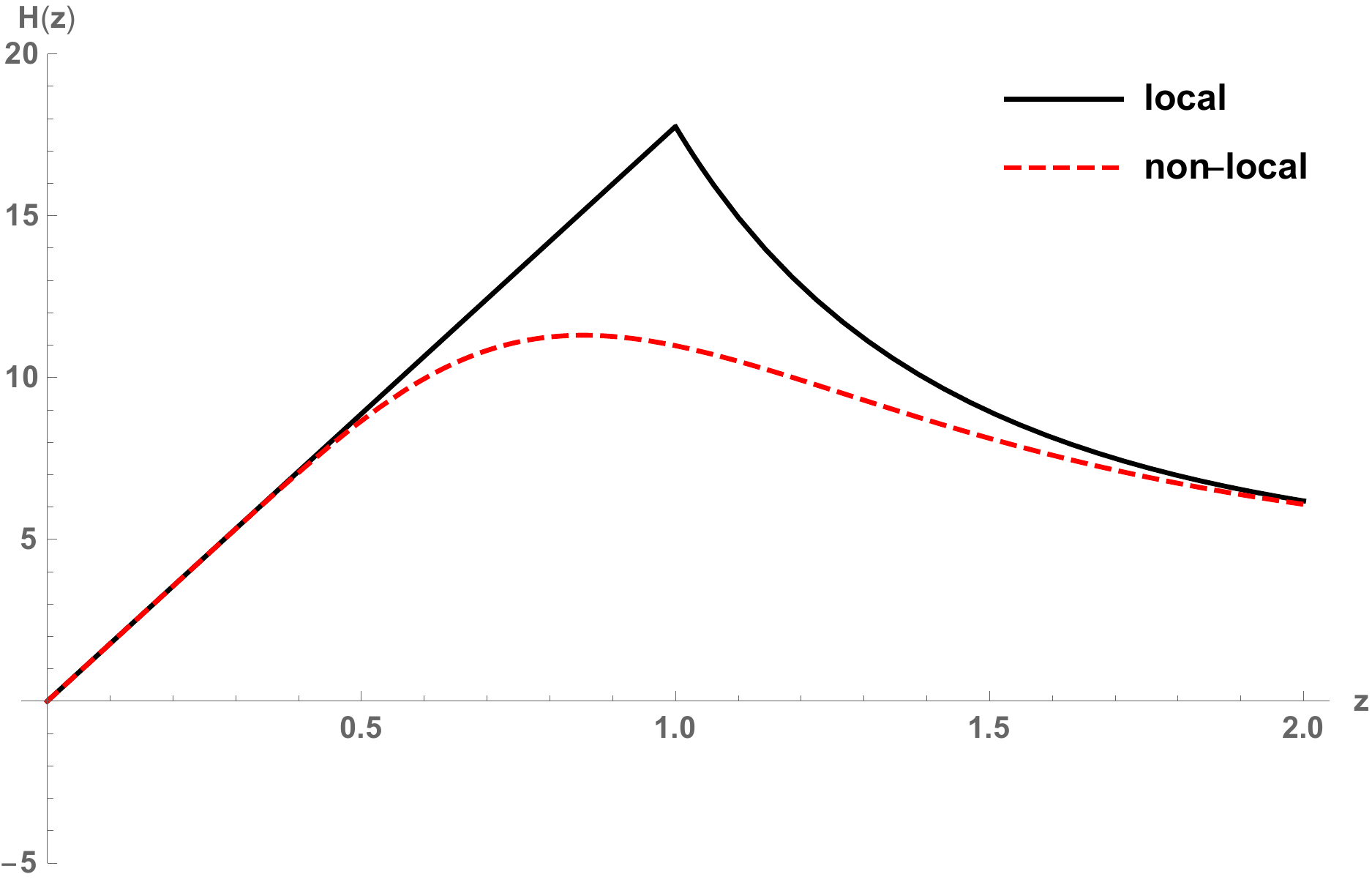}
	\caption{The function $H(z)$ for ${z_0=1}$, ${G=1}$, ${E=1}$, ${\ell=1}$, and ${M_s=4}$. The dashed red curve denotes the solution of IDG and the solid black curve represents the corresponding solution of GR.}
	\label{GraphSol4d2}
\end{figure}

This choice of the source allows the profile function $H$ to be independent of $y$. Thus, the field equation takes a simpler form
\begin{equation}
e^{-\frac{z^2\partial_z^2+2z\partial_{z}-2}{M_s^2\ell^2}} \left(z^2\partial_z^2+2z\partial_{z}-2\right)H(z)=-L_4\delta(z-z_0).
\label{eq:eqforlinsource}
\end{equation}
where ${L_4=16\pi G E z_0}$. Thanks to the absence of ${\partial_{y}}$, this equation can be solved directly using the \textit{heat-kernel method} \cite{Frolov:2015bia}. After transforming the equation to the coordinate ${w=\log{z}}$ and defining ${\tilde{H}(w)=H(e^w)}$, we can write
\begin{equation}
\begin{aligned}
\tilde{H}(w) &=-L_4 e^{-w_0}\frac{e^{(\partial_w^2+\partial_w-2)/{M_s^2\ell^2}}}{\partial_w^2+\partial_w-2}\delta(w-w_0)
\\
&=L_4 e^{-w_0}\!\!\int_{1/M_s^2\ell^2}^\infty\!\!\!\!\!\!\!\! ds\, e^{s(\partial_w^2+\partial_w-2)}\delta(w-w_0)
\\
&=L_4 e^{-w_0}\!\!\int_{1/M_s^2\ell^2}^\infty\!\!\!\!\!\!\!\! ds\, e^{-2s}e^{s\partial_w^2}\delta(w{-}w_0{+}s)
\\
&=L_4 e^{-w_0}\!\!\int_{1/M_s^2\ell^2}^\infty\!\!\!\!\!\!\!\! ds\, e^{-2s}\!\!\!\int_\mathbb{R}\!d\tilde{w}\,\frac{e^{-\frac{(w-\tilde{w})^2}{4s}}}{\sqrt{4\pi s}}\delta(\tilde{w}{-}w_0{+}s),
\end{aligned}
\end{equation}
where we applied the shift operator $e^{s\partial_w}$ on the third line and expressed the action of $e^{s\partial_w^2}$ using the heat kernel on the fourth line. This integral can be easily found. Returning back to the variable $z$, we obtain the particular solution of \eqref{eq:eqforlinsource},
\begin{equation}
\begin{aligned}
H(z)=\frac{8\pi GE }{3 z^{2}z_0}\bigg[& z_0^3\erfc\left(\frac{3}{2M_s\ell}-\frac{M_s\ell}{2}\log\Big(\frac{z}{z_0}\Big)\right)
\\
&+z^3 \erfc\left(\frac{3}{2M_s\ell}+\frac{M_s\ell}{2}\log\Big(\frac{z}{z_0}\Big)\right)\bigg],
\end{aligned}		
\end{equation}
which is plotted in Fig.~\ref{GraphSol4d2}.

By taking the local limit ${M_s\to\infty}$, we can recover the GR solution,
\begin{equation}
H_{\textrm{GR}}=\frac{8\pi GE  z}{3z_0}\bigg(1+\frac{z_0^3}{z^3}-\Big|1-\frac{z_0^3}{z^3}\Big|\bigg).
\end{equation}
As can be easily seen, this GR solution has a discontinuity and distributional curvature at the location of the source ${z=z_0}$, while the IDG solution is completely smooth everywhere. This is again caused by the fact that the form factor with infinite number of derivatives effectively smears the delta-like distributions in the stress-energy tensor. As before, the non-local solution approaches the GR solution near the conformal infinity ${z=0}$.

\section{Impulsive waves in 2+1 dimensions}\label{sc:SOL3D}

Now that we have discussed gravitational waves in ${3+1}$ dimensions, let us study the solutions in ${2+1}$ dimensions. In this section, we will not repeat details that remain almost the same, but focus on the important differences from the four-dimensional case. 

Since the Weyl tensor is identically zero in ${2+1}$ dimensions, the IDG action contains only the form factors of ${\cal F}_1(\square_s)$ and ${\cal F}_2(\square_s)$. Traceless part of the source-free field equations in three dimensions are reduced to
\begin{equation}
\begin{aligned}
&\bigg[1+\alpha_c\Big[-\frac{12}{\ell^2}\Big(f_{1,0}+\frac{f_{2,0}}{3}\Big)+\Big(\bar{\square}+\frac{2}{\ell^2} \Big){\cal F}_{2}(\bar{\square}_s)\Big]\bigg]\\&\times\bigg(\bar{\square}+\frac{2}{\ell^2}\bigg)\lambda_\mu\lambda_\nu H=0.
\end{aligned}
\end{equation}
Furthermore, one needs to set the form factor ${\cal F}_2(\square_s)$ to be in the following form in order to avoid ghost-like degrees of freedom \cite{Mazumdar:2018xjz}: 
\begin{equation}
\begin{aligned}
&{\cal F}_{2}(\square_s)=C\frac{e^{-(\square_s+\frac{2}{M_s^2\ell^2})}-1}
{\square_s+\frac{2}{M_s^2\ell^2}},
\label{formfactor}
\end{aligned}
\end{equation}
where we denoted ${C=1{+}\th(M_s^{-2}\ell^{-2})}$. It is also important to note that the field equation is independent of the form factor ${\cal F}_1(\square_s)$. We refer the reader to \cite{Mazumdar:2018xjz} for the explicit form of ${\cal F}_1(\square_s)$.

The AdS wave metric in ${2+1}$ dimensions is
\begin{equation}
ds^2=\frac{\ell^2}{z^2}\big(2du dv+dz^2\big)+2H(u,z) du^2.
\end{equation}
A similar arguments to those in Sec.~\ref{sc:APW} could be used to show that there are no new solutions of the homogeneous equation. In the next section, we focus on particular solutions in the presence of the non-zero source. 

\subsection{Massless point-like source}

Consider a point-like particle moving in the positive $x$-direction with the momentum ${p^\mu=E(\delta^\mu_t+\delta^\mu_x)}$ with the stress-energy tensor ${T_{uu}=E z_0^2\ell^{-2}\delta(u)\delta(z{-}z_0)}$. This source together with the impulsive-wave profile ${H=\delta(u)H(z)}$ lead to the equation 
\begin{equation}
e^{-\frac{z^2\partial_z^2+3z\partial_{z}}{M_s^2\ell^2}} \left(z^2\partial_z^2+3z\partial_{z}\right)H(z)=-L_3\delta(z-z_0),
\label{eq:eq2d}
\end{equation}
where ${L_3=16\pi G_3 E z_0^2/C}$. 

\begin{figure}[htbp]
	\centering
	\includegraphics[width=\columnwidth]{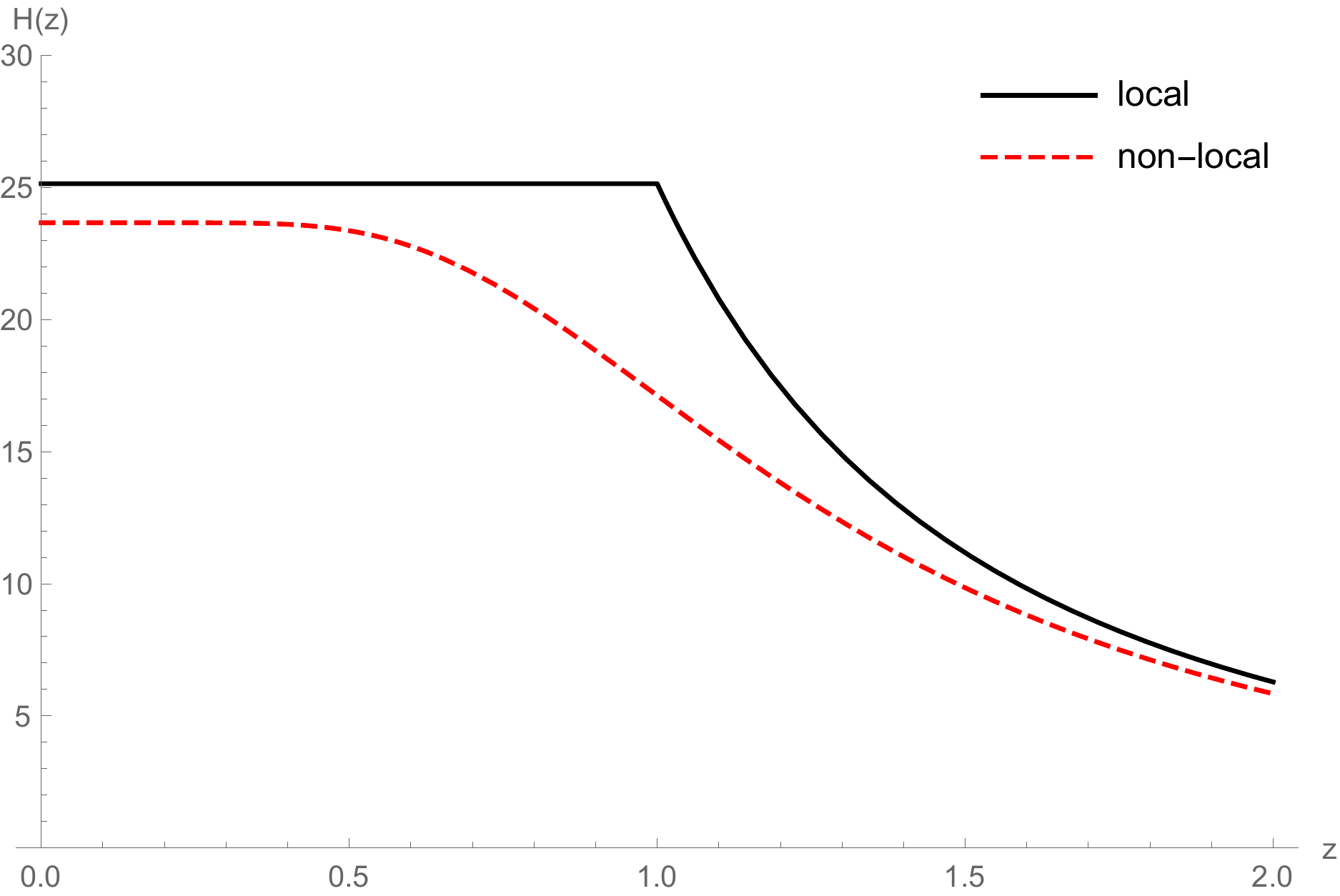}
	\caption{The function $H(z)$ for ${z_0=1}$, ${G_3=1}$, ${E=1}$, ${\ell=1}$, and ${M_s=4}$. The dashed red curve represents the solution of IDG and the solid black curve depicts the corresponding solution of GR.}
	\label{GraphSol3D1}
\end{figure}

By introducing ${w=\log{z}}$, ${\tilde{H}(w)=H(e^w)}$, and employing the heat-kernel method, we find
\begin{equation}
\begin{aligned}
\tilde{H}(w) &=-L_3 e^{-w_0}\frac{e^{(\partial_w^2+2\partial_w)/{M_s^2\ell^2}}}{\partial_w^2+2\partial_w}\delta(w-w_0)
\\
&=L_3 e^{-w_0}\!\!\int_{1/M_s^2\ell^2}^\infty\!\!\!\!\!\!\!\! ds \int_\mathbb{R}\!d\tilde{w}\,\frac{e^{-\frac{(w-\tilde{w})^2}{4s}}}{\sqrt{4\pi s}}\delta(\tilde{w}{-}w_0{+}2s),
\end{aligned}
\end{equation}
which can be easily calculated. The resulting particular solution of \eqref{eq:eq2d} is
\begin{equation}
\begin{aligned}
H(z)&=\frac{4\pi G_3E z_0 }{Cz^2}\bigg[z_0^2\erfc\left(\frac{1}{M_s\ell}-\frac{M_s\ell}{2}\log\Big(\frac{z}{z_0}\Big)\right)
\\
&\feq+z^2\erfc\left(\frac{1}{M_s\ell}+\frac{M_s\ell}{2}\log\Big(\frac{z}{z_0}\Big)\right)\bigg].
\label{exactshw3d}
\end{aligned}	
\end{equation}
This function is depicted in Fig.~\ref{GraphSol3D1}.

By calculating the local limit ${M_s\to\infty}$, we can arrive at the GR solution,
\begin{equation}
H_{\textrm{GR}}=4\pi G_3 E z_0\bigg(1+\frac{z_0^2}{z^2}-\Big|1-\frac{z_0^2}{z^2}\Big|\bigg).
\label{eq:3dsol}
\end{equation}
Unlike the four-dimensional GR solution of a point-like massless particle, which diverges, this three-dimensional GR solution is regular but has a discontinuity at ${z=z_0}$ \cite{Cai:1999dz}. This discontinuity is again cured by infinite derivatives. The IDG impulsive-wave solution is smooth everywhere. The full solution (with the homogeneous part ${c_1/z^2+c_2}$) approaches the GR solution at the conformal infinity ${z=0}$. Note that the metric of the GR solution is actually just the AdS metric. This is a consequence of the fact that the three-dimensional GR has no local degrees of freedom \cite{Deser:1983tn,Deser:1983nh}. The spacetime, however, differs from an empty AdS by a presence of non-trivial (global) topological defects that causes distributional curvature, which is not present in the non-local case.

\section{Conclusions}\label{sc:CON} 
In this paper, we studied the non-expanding gravitational waves of the Siklos type solutions of the ghost-free infinite derivative gravity in anti-de Sitter spacetime with the main focus on the impulsive waves which are generated by Dirac-delta source. We argued that the source-free infinite derivative gravity does not admit any new AdS wave solutions other than that of Einstein's general relativity. It was demonstrated that the non-locality described by form factors with the infinite number of derivatives plays a role only in the presence of a nonzero source.

We found the exact impulsive waves corresponding to massless point-like and linear sources propagating in four- and three-dimensional anti-de Sitter spacetimes. It turned out that the non-localities smear all the divergences and discontinuities (corresponding to distributional curvature) that are present in the local impulsive-wave solutions. The obtained solutions of the infinite derivative gravity are regular everywhere. They reduce to the impulsive-waves solutions of general relativity in the local limit ${M_s\to\infty}$ and in the infrared regime (near the conformal infinity of AdS). Simply put, the solutions get modified due to the non-local effects only in the ultraviolet regime, but not in the infrared regime.

\section{Acknowledgments}

We thank Bayram Tekin, Tahsin C. Sisman, Ayse K. Karasu, Pavel Krtou\v{s} and Eloy Ayon-Beato for useful discussions and suggestions. We also thank an anonymous referee for constructive suggestions. The works of S.D., E.K., and A.M. are supported by the TUBITAK Grant No. 119F241. I.K. and A.M. are supported by Netherlands Organization for Scientific Research (NWO) Grant number 680-91-119.

\appendix
\section{Equations of motion of IDG} \label{ap:EOM}
The equations of motion following from the action given \eqref{action} were found in \cite{Biswas:2013cha}. Using a common notation for a power of d'Alembert operator, ${\square^n X^{\alpha\dots}_{\beta\dots}\equiv X^{\alpha\dots}_{\beta\dots}{}^{(n)}}$, they can be written as
\begin{widetext}
\begin{equation}
\begin{aligned}
&G^{\alpha\beta}
{+}\Lambda g^{\alpha\beta}
{+}\frac{\alpha_c}{2}\Big[4G^{\alpha\beta}{\cal	F}_{1}(\square)R
{+}g^{\alpha\beta}R{\cal F}_1(\square)R
{-}4\left(\nabla^{\alpha}\nabla^{\beta}{-}g^{\alpha\beta}\square\right){\cal F}_{1}(\square)R
{-}2\Omega_{1}^{\alpha\beta}
{+}g^{\alpha\beta}(\Omega_{1}{}_\rho^{\rho}{+}\bar{\Omega}_{1}) 
{+}4R^{\alpha}{}_{\nu}{\cal F}_2(\square)R^{\nu\beta}
\\
&{-}g^{\alpha\beta}R_{\nu}{}^{\mu}{\cal F}_{2}(\square)R_{\mu}{}^{\nu}
{-}4\nabla_{\nu}\nabla^{\beta}({\cal F}_{2}(\square)R^{\nu\alpha})
{+}2\square({\cal F}_{2}(\square)R^{\alpha\beta})
{+}2g^{\alpha\beta}\nabla_{\mu}\nabla_{\nu}({\cal F}_{2}(\square)R^{\mu\nu})
{-}2\Omega_{2}^{\alpha\beta}
{+}g^{\alpha\beta}(\Omega_{2}{}^{\rho}_{\rho}{+}\bar{\Omega}_{2})
{-}4\Delta_{2}^{\alpha\beta}
\\
&{-}g^{\alpha\beta}C^{\mu\nu\rho\sigma}{\cal	F}_{3}(\square)C_{\mu\nu\rho\sigma}
{+}4C^{\alpha}{}_{\mu\nu\sigma} {\cal F}_{3}(\square)C^{\beta\mu\nu\sigma}
{-}4(R_{\mu\nu}+2\nabla_{\mu}\nabla_{\nu})({\cal F}_{3}(\square)C^{\beta\mu\nu\alpha})
{-}2\Omega_{3}^{\alpha\beta}
{+}g^{\alpha\beta}(\Omega_{3}{}_{\gamma}^{\gamma}{+}\bar{\Omega}_{3}) -8\Delta_{3}^{\alpha\beta}\Big]=0,
\label{IDGfeqns}
\end{aligned}
\end{equation}
where the symmetric tensors are
\begin{equation}
\begin{gathered}
\begin{aligned}
\Omega_{1}^{\alpha\beta} &=\sum_{n=1}^{\infty}f_{1,n}\sum_{l=0}^{n-1}\nabla^{\alpha}R^{(l)}\nabla^{\beta}R^{(n-l-1)},
&
\bar{\Omega}_{1} &=\sum_{n=1}^{\infty}f_{1,n}\sum_{l=0}^{n-1}R^{(l)}R^{(n-l)},
\\
\Omega_{2}^{\alpha\beta} &=\sum_{n=1}^{\infty}f_{2,n}\sum_{l=0}^{n-1}R_{\nu}{}^{\mu;\alpha(l)}R_{\mu}{}^{\nu;\beta(n-l-1)},
&
\bar{\Omega}_{2} &=\sum_{n=1}^{\infty}f_{2,n}\sum_{l=0}^{n-1}R_{\nu}{}^{\mu(l)}R_{\mu}{}^{\nu(n-l)},
\\
\Omega_{3}^{\alpha\beta} &=\sum_{n=1}^{\infty}f_{3,n}\sum_{l=0}^{n-1}C^{\mu}{}_{\nu\rho\sigma}^{;\alpha(l)}C_{\mu}{}^{\nu\rho\sigma;\beta(n-l-1)},
&
\bar{\Omega}_{3} &=\sum_{n=1}^{\infty}f_{3,n}\sum_{l=0}^{n-1}C^{\mu}{}_{\nu\rho\sigma}^{(l)}C_{\mu}{}^{\nu\rho\sigma(n-l)},
\end{aligned}
\\
\begin{aligned}
\Delta_{2}^{\alpha\beta} &=\frac{1}{2}\sum_{n=1}^{\infty}f_{2,n}\sum_{l=0}^{n-1}[R_{\sigma}{}^{\nu(l)}R^{(\beta|\sigma|;\alpha)(n-l-1)}-R_{\sigma}{}^{\nu;(\alpha(l)}R^{\beta)\sigma(n-l-1)}]_{;\nu},
\\
\Delta_{3}^{\alpha\beta} &=\frac{1}{2}\sum_{n=1}^{\infty}f_{3,n}\sum_{l=0}^{n-1}[C^{\rho\nu}{}_{\sigma\mu}^{(l)}C_{\rho}{}^{(\beta|\sigma\mu|;\alpha)(n-l-1)}-C^{\rho\nu}{}_{\sigma\mu}{}^{;(\alpha(l)}C_{\rho}{}^{\beta)\sigma\mu(n-l-1)}]_{;\nu}.
\end{aligned}
\end{gathered}
\end{equation}

\section{Poincar\'e spherical model} \label{ap:PSM}

\begin{figure*}[htpb]
	\centering
	\includegraphics[width=0.25\textwidth]{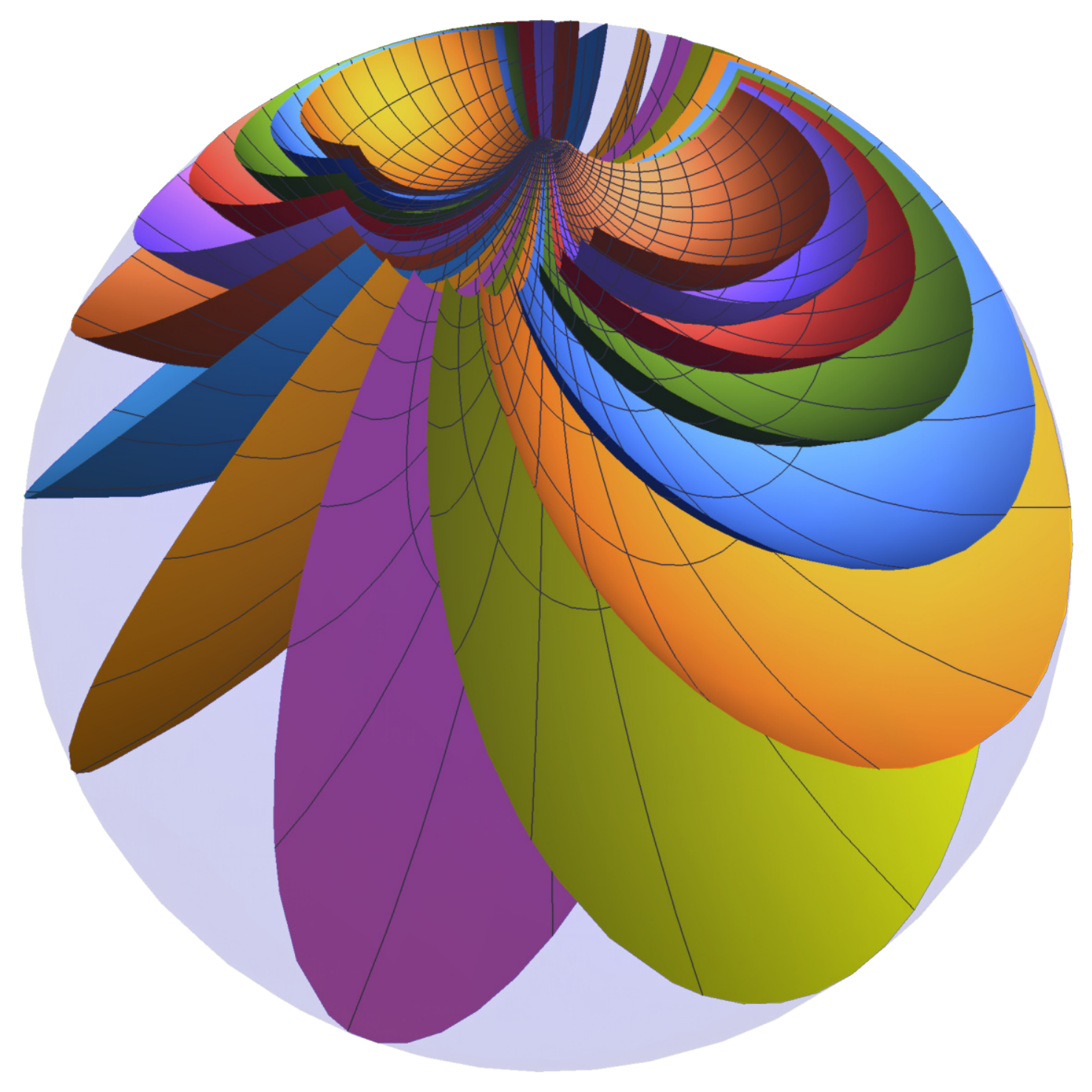}
	\includegraphics[width=0.25\textwidth]{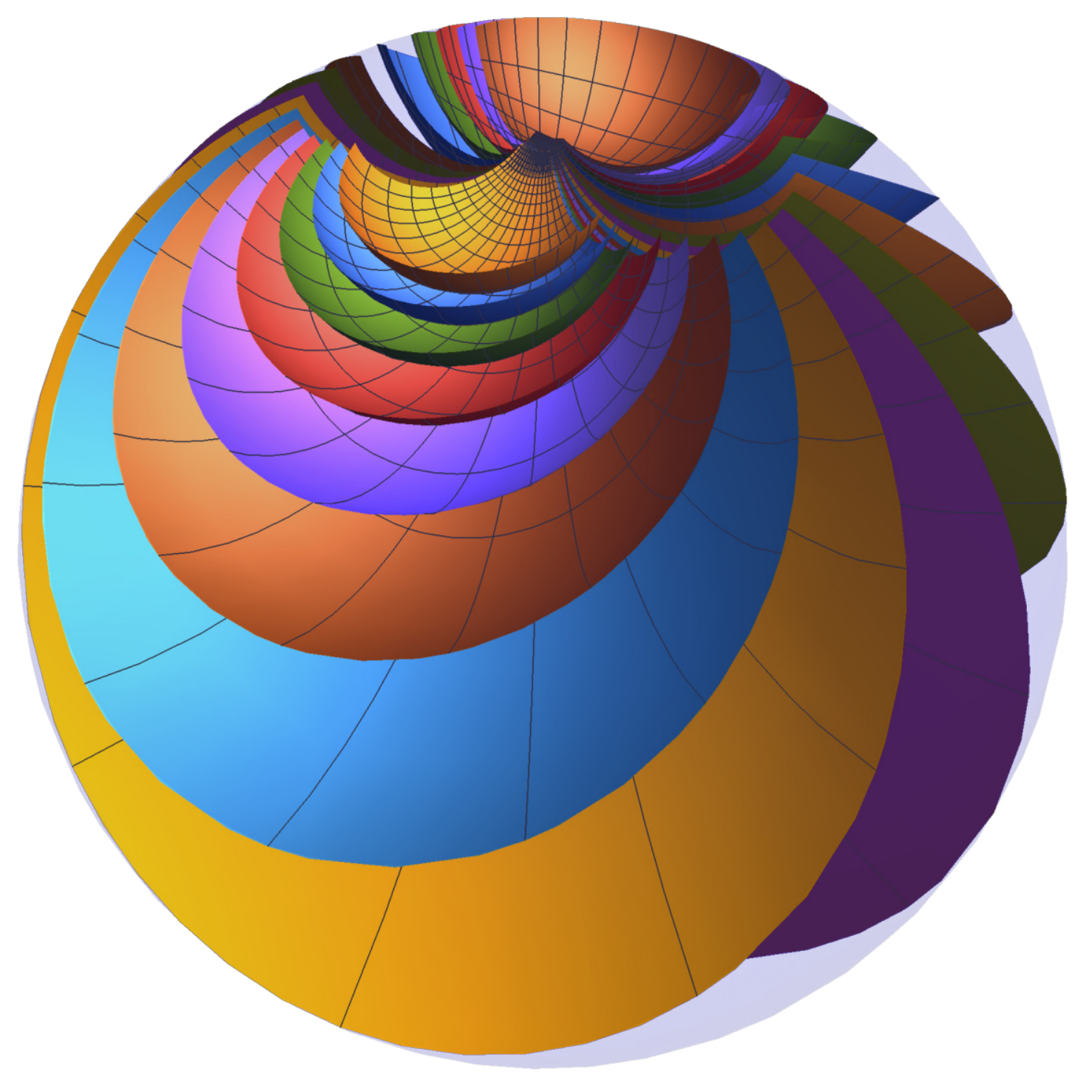}
	\includegraphics[width=0.25\textwidth]{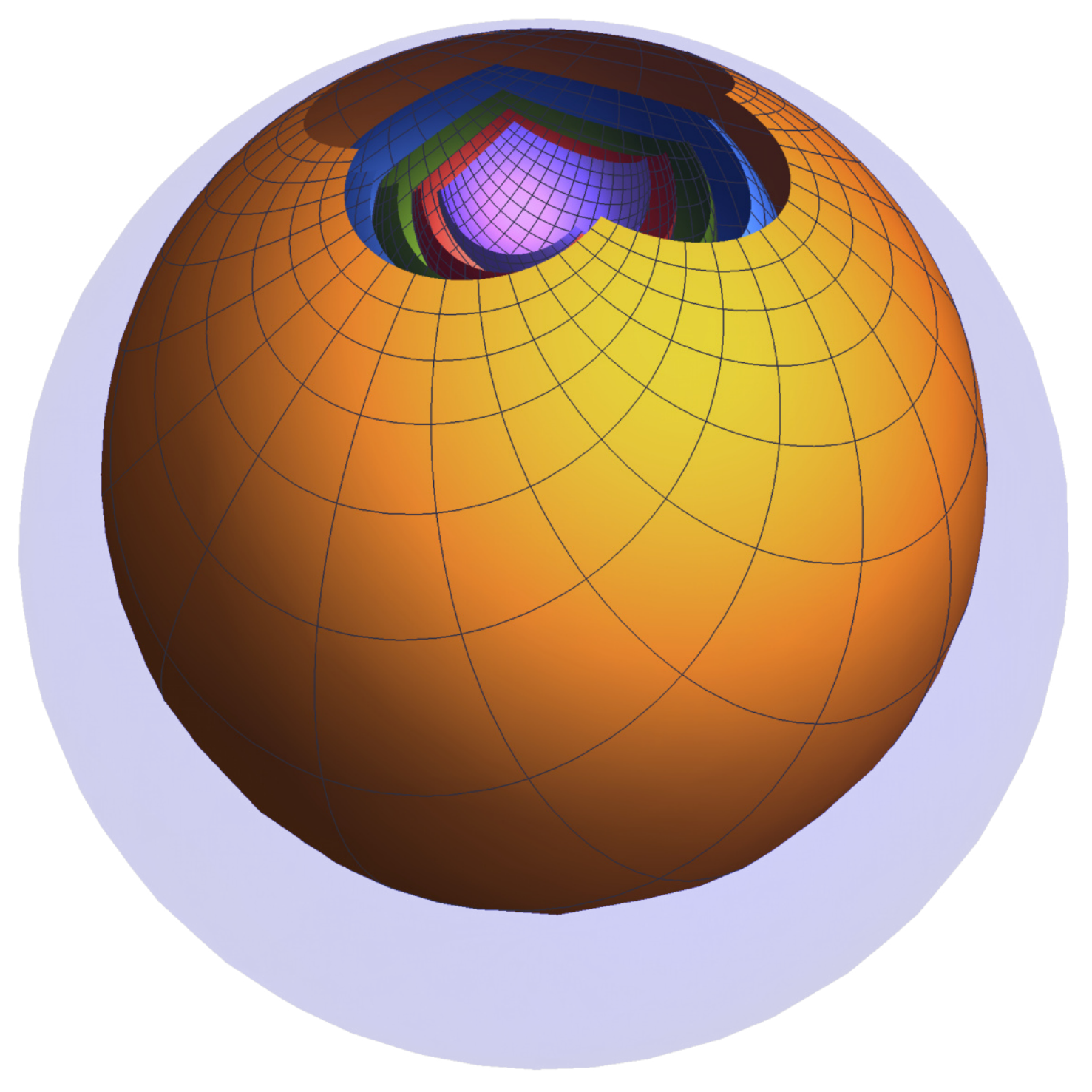}
	\caption{Surfaces of constant $x$, $y$, and $z$ coordinates of the Lobachevsky space depicted in the Poincar\'e spherical model.}
	\label{lobxyz}
\end{figure*}
Poincar\'e spherical model is a compactified representation of the Lobachevsky space,
\begin{equation}
ds^2=\frac{\ell^2}{z^2}\big(dx^2+dy^2+dz^2\big)\;,
\end{equation}
which is a spatial part of the AdS metric. The surface of the sphere is the conformal infinity. The standard conformally-flat coordinates $x$, $y$, and $z$ are visualized in Fig.~\ref{lobxyz}.

\end{widetext}

\end{document}